\documentclass[copyright,creativecommons]{eptcs}
\providecommand{\event}{}
\providecommand{\volume}{2}
\providecommand{\anno}{2009}
\providecommand{\firstpage}{99}
\usepackage{breakurl}
\usepackage[pdftex]{graphicx}
\usepackage{hyperref}

\title{Towards a Unifying View of QoS-Enhanced Web Service Description and Discovery Approaches}
\author{Dessislava Petrova-Antonova \qquad Sylvia Ilieva
\institute{Faculty of Mathematics and Informatics\\
University of Sofia ``St.~Kliment Ohridski''\\
Sofia, Bulgaria}
\email{d.petrova@fmi.uni-sofia.bg, sylvia@acad.bg}
}
\def\titlerunning{Towards a Unifying View of QoS-Enhanced WS Description and Discovery Approaches}
\def\authorrunning{D.~Petrova-Antonova \& S.~Ilieva}
\begin{document}
\maketitle

\begin{abstract}
The number of web services increased vastly in the last years. Various providers offer web services with the same functionality, so for web service consumers it is getting more complicated to select the web service, which best fits their requirements. That is why a lot of the research efforts point to discover semantic means for describing web services taking into account not only functional characteristics of services, but also the quality of service (QoS) properties such as availability, reliability, response time, trust, etc.\\
This motivated us to research current approaches presenting complete solutions for QoS enabled web service description, publication and discovery. In this paper we present comparative analysis of these approaches according to their common principals. Based on such analysis we extract the essential aspects from them and propose a pattern for the development of QoS-aware service-oriented architectures.
\end{abstract}
\section{Introduction}
The number of web services increased vastly in the last years. Various providers offer web services with the same functionality, so for web service consumers it is getting more complicated to select the web service, which best fits their requirements. That is why many research efforts point to discovering semantic approaches for describing web services including both functional and non-functional properties. This will give consumers the opportunity to find web services according to their QoS requirements such as availability, reliability, response time, trust, etc.\\
Most of the current solutions are based on Web Service Definition Language (WSDL) and Universal Description and Discovery Interface (UDDI) registry. WSDL documents provide functional description of web services without semantic specifications concerning QoS. UDDI registry provides catalog-based searching without control over the quality of registered services. UDDI APIs allow publishing and discovering data for a particular service, but do not provide an opportunity for a quality-based retrieval. The problem becomes more complicated when the discovery process returns several web services with the same functionality.\\
Such mentioned disadvantages motivated us to research principal current approaches for QoS-aware web service description and discovery in order to find better solutions giving more accurate and productive service retrieval. These approaches have common principals that can be summarized as follows:
\begin{itemize}
\item Ontologies and ontology languages used for description of functional and non-functional (QoS) characteristics of web services as well as for definition of clients requirements (WSDL, OWL-S, DAML-S, etc).
\item Repositories for storing web services descriptions such as UDDI registry, external databases, knowledge bases, semantic enhanced registry extensions, etc.
\item Classifications of QoS attributes used for definition of web service descriptions and clients requirements.
\item Description of strictly defined matching and ranking algorithms.
\item Collection of QoS information (sources of QoS could be QoS claims of providers, user feedback, agents, brokers, etc.).
\item Programming technologies used for implementation.
\end{itemize}
The paper presents a comparative analysis of those approaches that define the complete process of web service description, publication and discovery, which specially takes into account QoS. The main goal of the paper is to extract the essentials aspects from them and to summarize their characteristics as a pattern for development of QoS-aware service oriented architectures. Note that our research work is focused on the platforms and frameworks that implement all aspects of web service description and discovery process. Of course, there are many other approaches related to web service description and discovery problem, but they do not propose entire solutions. For example, Howard Foster et al. [25] propose a model-driven approach to dynamic and adaptive service brokering using Modes. The abstraction of modes is used for representation of changes in the service-oriented computing component architectures. The paper is focused on decentralized composition of services and applicability of extended modeling profiles to capture the dynamic change requirements anticipated for self-management. The problems, concerning QoS-aware web service description and discovery, are subject of the authors' research work within the Dino project [26].\\
The rest of the paper is organized as follows: Section 2 presents the mentioned approaches accordingly. Section 3 discusses the described in Section 2 approaches, extracts the main components that QoS-aware service oriented architectures should implement, and presents a pattern of QoS enhanced web service and discovery process. Section 4 concludes the paper and gives directions for future work.
\section{Web Service Description and Discovery Approaches}
Web service description and discovery process could be discussed from various aspects such as web service description language, services ranking criteria, trustworthy mechanism, etc. In this section, we will present how researched from us frameworks and platforms support this process relying on the aspects described in Section 1.
\subsection{Ontologies and Ontology Languages}
There is no preferred ontology or ontology language for offered and required service descriptions. WSDL is used in two cases. First, when the information about QoS attributes is stored in external database, for instance Quality database [18]; UDDI register is used to store only WSDL descriptions of functional web service characteristics. Second, in [18] the authors propose a modified WSDL for QoS-aware web service descriptions. The modified WSDL is the regular WSDL with more operations that correspond to the QoS information of the service.\\
One of the most usable ontology languages for semantic web service description is Web Ontology Language for Semantic Web Service Description (OWL-S). In [21], the offered and required serviced are specified through OWL-S or DAML-S profiles. DARPA Agent Mark-up Language for Services (DAML-S) is a semantic markup language for describing web services and related ontologies based on DAML+OIL. The DAML+OIL language uses earlier W3C standards such as RDF and RDF Schema, and ensure richer modeling primitives. In [11] a system that provides an automated web service client, augmented with semantic models, is proposed. OWL was chosen over RDF Schema as the ontology language for the semantic models. Pathal et al. [10] developed an approach for discovering Semantic Web services in which web service providers and web service requesters specify their advertisements and requests respectively using OWL-S.\\
For publishing, finding and searching web services Braun et al. [6] have chosen WSMO environment, which provides a concept for defining the functional aspects of services and requirements. The QoS upper ontology of the DIP project [1] is reused to describe the QoS properties. The service descriptions managed in the Web Service Repository are based on WSML. The upper ontology for context information called WSMO-Context is defined. Two types of context information are identified: measurable data and non-measurable data. The authors also present another orthogonal classification of context information based on the differentiation between static and dynamic context. Wang et al. [5] also use WSMO to describe QoS model, specific quality metrics, value attributes, and their respective measurements. They specify a QoS upper ontology named WSMO-QoS. It is a complementary ontology that provides detailed quality aspects about services. Each QoS metric is generally described by MetricName, ValueType, Value (given or calculated at service run-time), MeasurementUnits (e.g.~\$, millisecond), ValueDefinition (how to calculate the value of this metric), and Dynamic/Static. The WSMO conceptual model is also applied from Vu et al. [1] in order to describe web services semantically, including QoS properties.\\
Corrales et al. [17] have developed the BeMatch platform for ranking web services based on behavior matchmaking. They present service behavior using Business Process Execution Language (BPEL) and Web Services Conversation Language (WSCL).\\
In [3] the Registry Ontology is created for describing the semantic of QoS Information and Computation framework, the semantic of Web services and their associated QoS properties. QoS ontology is established to overcome the ontological conflicts that may occur between consumers and providers. It defines the QoS properties, their relationships and establishes shared conceptions between consumers and providers.\\
Maximilien and Singh propose an approach [13] that implements an agent-based architecture based on SOA that is realized in the Web Service Agent Framework (WSAF). They have defined QoS ontology that lets service agents match advertised quality levels for its consumers with specific QoS preferences. The upper ontology captures the most generic quality concepts and defines the basic concepts associated with a quality, such as quality measurement and relationships. The QoS middle ontology describes the following QoS attributes: availability, capacity, Economic conditions, interoperability, performance, reliability, robustness, scalability, security, integrity and stability.
\subsection{Repositories for Storing Web Services Descriptions}
The most of the proposed frameworks and architectures use UDDI registry for storing web service descriptions. As was mentioned above the UDDI register can be supported with external database, which stores non-functional information about web services. For instance, the proposed in [1] architecture uses QoS data repository to store the feedback of QoS reporters, the reputation information of all service providers as well as the evaluated QoS information of QoS-aware web services. In addition, different extensions of UDDI registers are designed to store QoS characteristics.\\
Kourtesis and Paraskakis [20] describe the implementation of the FUSION Semantic Registry, a semantically enhanced service registry, developed within the FUSION project and released as open source software. The FUSION Semantic Registry exposes two specialized Web service APIs to the client for publication and discovery functions. It is also responsible for performing the associated SAWSDL parsing, OWL ontology processing and DL reasoning operations. It consists of UDDI, OWL KB module, Publication manager and Discovery manager. UDDI server stands independently to the semantically-enabled service registry modules in the proposed architecture. The OWL KB module is a typical OWL ontology with RDF/XML serialization that the Semantic Registry uses for storing the Advertisement Functional Profiles it generates at the time of service publication. The Publication Manager module provides a Web service interface to the user for adding, removing, or updating Web service advertisements, as well as adding, removing, or updating descriptions of service providers. The Discovery Manager module provides a Web service interface for retrieving a specific service advertisement or service provider record via its key, discovering a set of services or service providers through keyword-based for terms contained in their names, and most importantly, discovering a set of services based on a Request Functional Profile.\\
ShaikhAli et al. [16] extend UDDI as 'UDDIe' in order to avoid some of the UDDI restrictions. Extensions in the UDDIe are based on four types of information: business information; service information, binding information; and information about specifications for services. In the UDDIe a businessService structure represents a logical service---and it is the logical child of a businessEntity---the provider of the service. Service properties are part of the propertyBag entities---such as QoS that a service can provide, or the methods available within a service that can be called by other services. It is an important feature missing in current UDDI implementations. The API for interacting with the registry system extends three classes within existing UDDI implementations. The extensions provided in the API include: saveService, findService, getServiceDetails, renewLease, startLeaseManager. In addition to these sets of APIs, the authors introduce support for a Qualifier-based search.\\
In [9] the authors propose Enhanced-UDDI register, essentially a layer above UDDI, which is capable of handling semantic data. In the presented implementation the value attribute of the UDDI category bag's elements is used to hold the semantic content. Binding templates hold Location and Domain specific tModels. This enables direct search of services that function in a particular Geographic Location and Domain. The category bag associated with the Business Service, serves as a placeholder for the operation /inputs /outputs /exceptions /constraints oriented semantics. Service specific semantic information is stored in the Binding Template, which falls under Business Service. This abstraction of data helps to organize the information for effective retrieval during discovery. An advertisement built from the annotated source code semantic descriptions serves as the input to the Publishing interface. The discovery Engine employs a query similar to the advertisement for finding the information from the Enhanced-UDDI.\\
Ran [12] proposes a new service discovery model that can co-exist with the current UDDI registers where quality of service is taken as constraint when searching for Web services. There are four roles defined in that model: Web service supplier, Web service consumer, Web service QoS certifier, and the new UDDI registry. The Web service provider offers Web service by publishing the service into the registry; the Web service consumer needs the Web service offered by the provider; the new UDDI registry is a repository of registered Web services with lookup facilities; the new certifier's role is to verify service provider's QoS claims. UDDI registry is enhanced with new data structure named 'qualityInformation'. It represents description of QoS information (organized into four categories described in section 2.3) about a particular service and refers to tModels defined in the UDDI registry.\\
In [8] authors describe UDDI eXtention (UI) that facilitates requesters to discover services with good qualities. The extended inquiry interface conforms to the UDDI specification. Additional policies are recommended to manage behavior of the registry.\\
Taher et al. [3] have designed aeRegistry that is composed of four new internal components beside the UDDI registry. The new internal components are following: SOAP Request Filterer (receives lookup/publish requests form customers/providers), Update Manager (handles requests from providers to update the service registry with changes in QoS properties and actual consumers of published services), QoS Manager (updates QoS properties based on data received from Update Manager and provide QoS selection algorithm), Validation Manager (validate taxonomy and business information submitted by providers).
\subsection{Classifications of QoS Attributes}
Currently QoS characteristics of web services are not defined in a standardized manner. Various authors categorized them in different ways.\\
Patel et al. [15] develop a QoS-based Web Service framework, called WebQ that enables to select appropriate Web Services, dynamically bind the services with the underlying workflow, and perform the refinement of existing services. They classify QoS parameters into the following three categories: General (Latency, Throughput, Reliability, and Cost), Internet Service Specific (Availability, Security, Accessibility, and Regulatory) and Task Specific (related to the quality of the output or the type of service offered etc.).\\
Liu et al. [7] propose extensible QoS model including generic and domain or business specific criteria. Three generic quality criteria, which can be measured objectively for elementary services, are considered: execution price, execution duration and reputation. The number of business related criteria can vary in different domains. In the current application, usability is measured from three aspects, transaction, compensation rate and the penalty rate. Some authors, for instance in [8, 11], observe only generic QoS attributes such as availability, reliability, execution/response time and cost.\\
In [2] the set of service features is composed by aggregating two types of attributes: Objective attributes (e.g.~response time), which can be measured and evaluated exactly, and Subjective attributes that express the user opinion regarding the web service from a specific point of view (ex. usability).\\
In [12] the QoS attributes are organized into four categories: runtime related QoS (scalability, capacity, performance, reliability, availability, robustness/flexibility, exception handling, accuracy), transaction support related QoS (integrity), configuration management and cost related QoS (regulatory, supported standard, stability/change cycle, guaranteed messaging requirements, cost, completeness) and security related (authentication, authorization, confidentiality, accountability, traceability and auditability, data encryption, non-repudiation).
\subsection{Matching and Ranking Algorithms}
The matchmaking and ranking algorithms have a key role in the process of web service discovery. The result of the matching algorithm is a list of all services which satisfy the functional or both functional and QoS requirements of the client. The result set of candidate web services would be ranked based on the degree of match [22] or client ranking criteria [10]. Thus the client can invoke the best web service.\\
The matching algorithm proposed in [21] consists of four faces. It extends the algorithm of Paolucci et al [22]. All advertised services that belong to the same category and requested by the client are filtered during the first phase. The second phase consists in establishing the different possible combinations between the request given by the client and the advertisement published by the providers that due to the previous phase belong to the same service category. During the third phase each time a couple is obtained in the previous phase, different degrees of similarity for functional parameters must be applied to them, and the relative weights must be calculated for functional parameters and non-functional parameters. The last phase takes into account the weights calculated and executes the ordering algorithm on the service list. The service that heads the list will be one that is considered optimal. The rest of the list will be maintained for fault tolerance.\\
The authors proposed seven degree of matching for functional parameters. The matching algorithm searches for exact coincidence in non-functional parameters. It uses four filters: region filter, quality service filter, service name filter and provider name filter. Region filter checks if the geographical ratio given by the client is the same as required from the provider. Quality service filter consists in checking in the repository whether the quality provided by Web services is the same as that requested by the client. A ranking for quality of traffic services was made. Service name filter checks whether the client finds any service in particular with the name provided. This matching is syntactic, since the way in which the service-name profile is defined in the Profile subontology has as a value range, the String datatype from XML Schema, with which only syntactic comparisons can be made. Provider name filter allows checking whether the Web service is provided by the company in which the client is interested. As the previous filter only a syntactic match is made since this Profile property is defined as an XML Schema String. The weight of these parameters is globally calculated, although they affect in different ways depending on their nature (syntax or semantics).\\
The discovery query that initiates the semantic matchmaking process in [20] comprises two elements: (i) a URI pointing to some Request Functional Profile (RFP), and (ii) an optional UUID designating the preferred service provider. The RFP that the URI points to may be defined within an ontology that is shared by service providers and service requestors alike (i.e.~be a reusable RFP defined in the FUSION Ontology), or within some third-party ontology that imports and extends the shared ontology (i.e.~be a custom-built and non-shared RFP). Depending on which of the two cases holds, the algorithm would follow a different discovery path. The result of the discovery process, regardless of the ontology in which the RFP is defined, is a list of UUID keys corresponding to advertisements of services that comply with the matchmaking criteria modeled in the RFP. If a service provider UUID has been also specified in the discovery query, the UDDI server will restrict the result set to only those services offered by the specified provider.\\
Makris et al.~\cite{19} propose an efficient and adaptive algorithm that performs selection among similar Web Services located at different infrastructures. The algorithm includes two functional components that compose the final step of the Web Service selection---Contour Selection component and Adaptive Selection component. Contour Selection component takes into consideration two parameters and based on them, it performs an efficient selection of the best candidate Web Services. The parameters include (i) the network distance (network latency between the client requester and the Web Service), (ii) the number of other distinct Web Services, functionally related to the Web Service in terms of business environment. The second component presents adaptive selection process. It is based on online quality ratings of a Web Service (QoWS ratings). These are measures, i.e., countable factors such as available memory, etc., concerning the quality and the availability conditions of the infrastructure implementing the WS at the very execution moment. A service's average execution time is observed over a number of executions over time. At this point to assure the maximization of quality in the chosen Web Service, the fast corporate network is utilized to collect up-to-date WS measurements and deduce the estimation of the expected QoWS characteristics based on previously recorded WS execution schemes-profiles.\\
In~\cite{17} authors propose behavior matchmaking. The first step for selecting possible candidates is based on non-operational information. This discovery step is implemented by means of a hierarchically structured catalogue. After the first discovery step, a list of candidate services is found. Then, the query service is compared with each candidate service and the ranked list of candidates is presented to the requestor. The service ranking is based on service behavioral matching which is reduced to a graph matching problem. The authors use the error correcting subgraph isomorphism (ECSI) algorithm in order to allow an approximate matching. The data in the repository is organized in a hierarchical manner. The services are filtered in order to reduce the number of candidates for the second step of the discovery process, the behavioral matchmaking. Services to graphs parser transform a service behavior description to a graph. The Graph matchmaking module takes as inputs the query graph and a single target graph (from the candidates list) produced by the parser presented above and computes the semantic distance between them by applying the algorithm for optimal error correcting subgraph isomorphism detection. The module of Cost functions groups the cost functions for the graph edit operations that allow calculating the distance between graphs. The costs assigned to different graph edit operations reflect the relative importance of dissimilarities between different graph attributes. Thus they depend of service behavior metamodel and on the application domain. The Linguistic analyzer calculates the linguistic similarity between two strings using the following algorithms: NGram, Check synonym, Check abbreviation and tokenization. Granularity level analyzer checks whether decomposition/ composition operations are necessary and add their cost to the graph edit distance. The module of Similarity functions defines the similarity functions that allow constructing the service ranking. The Tool for evaluating the effectiveness of the behavioral matchmaking method allows creating a service ranking based on manual comparisons between a query service and a list of target services in the repository. The tool permits to compare the ranking defined by the user with the results of a matchmaking tool. The cost functions for graph edit operations use some user defined parameters (weights reflecting the importance of different service attributes in evaluating the similarity). Given the fact that the parameterization of the cost function is domain dependent and very important for the effectiveness of the matchmaking method, this tool helps the user determine the optimal parameters to apply for a given domain and similarity criteria.\\
The selection algorithm used in~\cite{16} is based on the Weighted Average (WA) concept. In this algorithm, the authors introduce the notion of a QoS importance level, whereby the client/application is requested to associate a level of importance, such as High, Medium or Low, with every QoS attribute. Based on this QoS importance level and the value of a QoS attribute, the algorithm computes the WA for every returned service and selects the service with the highest WA.\\
The service selection criteria in the framework proposed in~\cite{10} comprises of two components: Selection of the service providers and then, Ranking the selected providers. The first step in service selection is to determine a set of service providers which offer the requested functionality. The services are categorized based on the degree of match: Exact, Plug-in, Subsumption, Intersection, and Disjoint. An implicit ranking amongst the potential providers based on the service categorization is provided. Only the services which belong to Exact, Plug-in and Subsumption are considered. The second step further refines the set of candidate service providers based on user-specified QoS attributes. These attributes are used to compose a quality vector comprising of their values for each candidate service. Quality vectors form a quality matrix such that, each row of the matrix corresponds to the value of a particular QoS attribute and each column refers to a particular candidate service. Quality matrix is used to select one or more services, which satisfies user's constraints and has an overall score (for the non-functional attributes) greater than some threshold value specified by the user. If several services satisfy these constraints, then they would be ranked according to the user-specified ranking criteria. But, if no service exists, then an exception is raised and the user is notified appropriately. The authors propose a notion of ranking attributes and a ranking function, which will be used to rank the selected candidate service providers. Once the service providers are ranked, it is left at user's discretion to select the most suitable provider.\\
In~\cite{5} the matrix of QoS is used for service matchmaking. The quality requirements are in the first row of the matrix and the quality information of candidate services is in the other rows. Each column contains values of the same quality property. The main idea of the algorithm is to scale the value ranges with the maximum and minimum values of each quality metric for thousands of current candidate services. Accordingly, the maximum and minimum values are mapped to the uniform values 1 and 0, respectively, depending totally on their definition of hasTendency. The second pre-processing step is uniformity analysis. Different quality metrics with their value features are distinguished. In the proposed QoS model, the information of hasTendency is taken as a quality metric. The weighted value for each quality metric is defined in the parameter of hasWeight. Finally, the evaluation result for each quality metric is calculated by summing the values of each row. These abstract values are taken as a relative evaluation of each service's QoS.\\
The proposed framework in~\cite{3} uses standard UDDI registry extended with four internal components: SOAP Request Filterer, Update Manger, QoS Manager and Validation Manager. For the purpose of matching QoS Manager constructs matrix. Each row in the matrix represents a Web service, while each column represents one of the QoS properties. The authors use Euclidean distance to calculate the distance between the users specified QoS properties and the existing QoS properties for each vector in the matrix. Then the algorithm finds the web service with the minimum Euclidean distance. In the matching algorithm weights to QoS parameters to reflect consumer's preferences on specific QoS properties are missing.
\subsection{Collection of QoS Information and Trustworthy}
Different techniques for collection of QoS information are presented in this section. Some of them use advertisements of web service providers and feedback from clients for QoS information retrieval. Other use third parties---like agents, monitors or brokers---to collect information, which makes proposed architectures more trustworthy.\\
WebQ framework presented in~\cite{15} uses a monitor, which is responsible for monitoring, measuring and asserting facts about the QoS parameters. The facts are asserted in a JESS Knowledge base depending on the rules fired. Knowledge Base is a repository of facts about Web QoS parameters. Rule Repository collects rules used to specify user specific QoS requirements, available workflow QoS and elicitation of steps to be taken for achieving specified QoS requirement. Jess Knowledge base is used also in~\cite{10}. It stores a collection of JESS facts, which is obtained at the time of service advertisements registration.\\
Chukmol et al.~\cite{14} propose a collaborative tagging-based environment for Web service discovery, allowing users to tag or annotate a Web service using keyword or free-text. Before tagging a resource (referring to a Web service), it should be used or tested so that users can judge on its functionality and other related qualitative aspects.\\
For each service, WSAF~\cite{13} creates a service agent that exposes the service's interface, augmented with functionality to capture the consumer's QoS preferences or policies and to query agencies or other agents for a suitable match. The agent can determine objective QoS-attribute values on its own and get user feedback for subjective attributes. It then conveys these QoS values to the appropriate agencies.\\
In the model presented in~\cite{12} the web service provider announces QoS characteristics of a particular web service. The claimed QoS needs to be certified and registered in the repository. That is why a special certifier checks the QoS claims and issues certification identified by a certification ID. After that the provider can register its service with both functional description and associated certified QoS information.\\
The proposed by Julian Day system~\cite{11} consist of two parts: augmented client and the QoS forums. The clients send their experiences to a central web service which stores this information inside an internal database. This web service can be thought of as a kind of forum system for QoS information. It can respond to requests about particular web services, sending all the data it knows about a particular service to a requesting client. Now when a client wants to pick a service, it gathers information from the QoS forums, and then reasons about which service is best.\\
METEOR-S~\cite{9} is a front-end tool for source code annotation and semantic Web service description generation. It has Semantic Web Service Designer, which is a GUI to design and develop Semantic Web Services. Using this tool, interface design of services and incorporation of semantic description into the service can be developed simultaneously. The output of the Semantic Web Service Designer is an annotated source code. Semantic Description Generator generates annotated WSDL 1.1 files and WSDL-S files, which provide web service descriptions. It also generates OWL-S files associated with the annotated source code. OWL-S files provide a more complex representation of the semantic descriptions.\\
UX system architecture proposed in~\cite{8} is comprised of service requester, local UDDI registry, Test host and UX Server. The test host is designed to generate QoS reports for the services registered in local registry. It tests the service with random or pre-defined parameters to gain service reports.\\
Service quality information in~\cite{7} can be collected via active monitoring and user's feedback. Through active execution monitoring, when a service requester gets a different set of values for the deterministic criteria advertised by the service provider, this difference can be logged. If the difference is bigger, the lower QoS value is assigned to the relevant service provider. Each end-user is required to update QoS of service he or she has just consumed. To prevent the manipulation of QoS by a single party, for each feedback, the end-user is given a pair of keys. This pair of keys must be authenticated by the service provider before the user is allowed to update the QoS value.\\
Letia and Pop propose a mechanism~\cite{2} for web services selection based on customers' review and feedback. Their approach is based on the existence of a broker between service client and web service provider. The broker is an independent component that will rate truthfully web service usages. So, the algorithm for estimating trust gathers reviews from users, as a subjective feedback, and data provided by the broker as objectively collected data. The trust problem is also referred by Vu et al. They propose reputation-based model~\cite{1}, which exploits data not only from user's feedback but from QoS values promised by providers in their service advertisements, reports produced by a few trustworthy QoS monitoring agents and different discovery components in the network that periodically exchange reputation information.
\subsection{Programming Technologies used for Implementation}
Since the approaches discussed in this paper propose complete frameworks for web service description and discovery, most of them have concrete implementation with appropriate programming technologies.\\
The proposed approach in~\cite{18} is implemented using the .NET v.1.1 framework technology and C\# language. A Quality database is designed and runs on an MS SQL Server 2000 sp4 instance. For the UDDI registry connection the MS online UDDI catalogue available for testing purposes has been utilized at Microsoft UDDI Business Registry Node-Test Site. .NET v.1.1 framework technology and C\# language are also used by authors in~\cite{19}.\\
The platform in~\cite{17} is built using the Java programming language (JDK 1.6.0) and the Netbeans 5.0 Integrated Development Environment (IDE). The desktop implementation uses the Xerces Application Program Interface (API) for processing the BPEL and WSCL documents within the Parser Module. The JGraph and JGraphLayout APIs were used for the visual representation of service models. The Linguistic analyzer accesses the Wordnet Dictionary through the Java Word Net Library (JWNL) API for verifying the semantic relationship between two words.\\
The authors in~\cite{15} have implemented and deployed a Java based prototype of the proposed framework on Linux. The Java Web Services toolkit (JAX-RPC) is used for implementing the in-house Web Services and for interacting with third party Web Services. The first prototype of the proposal in~\cite{14} is being also implemented as a Java Web based application.\\
A prototype of the local UX system~\cite{8} is implemented using Apache Axis, UDDI4J and WSIF as basic components. The IBM UDDI registry software is used as a local registry. The presented prototype in~\cite{6} is entirely written in Java. SOAP/HTTP is used as a protocol for the communication between the components. For visualization, JSP-Clients are implemented for both interfaces. The implementation of the framework described in~\cite{3} is based on jUDDI registry, Tomcat application server, and Apache Axis and MYSQL database.
\section{Discussion}
As it can be seen from Table 1 most of the presented in Section 2 frameworks use semantic ontology languages in order to enrich service descriptions with QoS. This is due to the big variety of ontologies and ontology languages for semantic resource description developed in the recent years. The authors who use WSDL store QoS information in external repositories such as Quality databases or Knowledge bases because so far there is no language describing both functional and non-functional service characteristics that could be incorporated in the current UDDI registries. That is why many research efforts concerning web services are concentrated on design and development of UDDI extensions or entirely new semantic enhanced repositories. Section 2.2 presents detailed information on this topic.\\
One of the objectives of our research work is to determine the most important QoS properties in the process of service description and discovery. Still there is no widely accepted QoS standard, which specifies QoS attributes as well as their measures and monitoring contexts. QoS aspects and desired requirements are described in the first W3C Note of the Web Services QoS Requirements and Possible Approaches document. This document includes definition of 13 QoS properties: performance, reliability, scalability, capacity, robustness, exception handling, accuracy, integrity, accessibility, availability, interoperability, security, and network-related QoS requirements~\cite{23}. Most of them are taken into consideration in the frameworks and platforms described in Section 3, but there is not enough information on how exactly they could be monitored and measured during the web services' execution. QoS metrics are often described in semantic ontologies for web service descriptions. For instance, DAML-QoS ontology has QoS Metrics Layer providing the QoS metrics definition for the QoS property's range constraints. It specifies precise semantic meanings for service measurement partner to measure the service and check against the guarantee. QoS properties are specified in Web Services Quality Model (WSQM) of OASIS. QoS properties, named Quality sub-factors in the model, include Response Time, Throughput, and Maximum Throughput and Stability related quality sub-factors include Availability, Reliability, and Accessibility~\cite{24}. Each Quality sub-factor is presented with its definition and calculated formula. In this way, the first steps in the creation of common web service Quality standard are already done.\\
The essential part of the proposed architectures are matching and ranking algorithms. They collect QoS information not only from services descriptions advertised from providers, but also from client experience in the form of feedback. Third parties such as brokers or agents are also included in the discovery process. They provide QoS reports based on monitoring and testing. Thereby, the discovery process becomes more reliable and contributes to solving the problems such as those in trust and reputation systems.\\
Table 1 shows that all proposed approaches solve the problem concerning Quality-aware web service description from ontology language and web service repository point of view. About 70\% of them use semantic enhanced ontology language and repository, which is an extension of UDDI registry or an entirely new solution. As we mentioned above some of the authors describe QoS properties in their publications (about 30\%). Others leave this problem open or rely on the definitions in an existing ontology. About 50\% of them describe the ways that they use to collect QoS information and specify web service selection algorithms. The last column of the table shows that approximately 30\% of approaches are implemented with using of concrete programming technologies and languages. Existing implementations prove the depth of the current research results because each implementation needs detailed specification of web services description and discovery process including all the aspects discussed above.
\begin{table}
\caption{Web service description and discovery aspects in different QoS-aware frameworks.}
\small
\centering
\begin{tabular}{|l|p{2.1cm}|p{.9cm}|p{1.2cm}|p{.8cm}|p{1.3cm}|p{1cm}|p{1.4cm}|p{1cm}|p{.9cm}|}
\hline
\textbf{No} & Name of the\break first author of\break the publication & Use\break of\break WSDL & Use of\break semantic\break ontology\break language & Use\break of\break UDDI & Use of\break extended\break UDDI or\break external\break repository & Def.~of\break QoS\break proper- ties & Def.~of\break matching\break and\break ranking\break algorithms & Def.~of\break QoS\break sources & Imple- men-\break tation\\
\hline
\hline
1 & Vu &  & X &  & X &  &  & X & \\
\hline
2 & Letia &  &  &  &  & X &  & X & \\
\hline
3 & Taher &  & X &  & X &  & X &  & X\\
\hline
4 & Kokash & X &  & X &  &  &  &  & \\
\hline
5 & Wang &  & X &  &  &  & X &  & \\
\hline
6 & Braun &  & X &  & X &  &  &  & X\\
\hline
7 & Liu & X &  &  & X & X &  & X & \\
\hline
8 & Zhou & X &  &  & X & X &  & X & X\\
\hline
9 & Rajasekaran & X &  &  & X &  &  & X & \\
\hline
10 & Pathak &  & X &  & X &  & X & X & \\
\hline
11 & Day &  & X &  & X & X &  & X & \\
\hline
12 & Ran & X &  &  & X & X &  & X & \\
\hline
13 & Maximilien &  & X & X &  &  &  & X & \\
\hline
14 & Chukmol & X &  & X &  &  & X & X & X\\
\hline
15 & Patel &  & X & X &  & X &  & X & X\\
\hline
16 & ShaikhAli & X &  &  & X &  & X &  & \\
\hline
17 & Corrales &  & X &  & X &  & X &  & X\\
\hline
18 & Diamadopoulou &  & X & X &  &  &  &  & X\\
\hline
19 & Makris &  &  & X &  &  & X &  & X\\
\hline
20 & Kourtesis &  & X &  & X &  & X &  & \\
\hline
21 & Samper &  & X &  & X &  & X &  & \\
\hline
\end{tabular}
\end{table}\\
Based on our research work, we can summarize that each QoS-aware service oriented architecture should have the following main components showed on Figure 1:
\begin{itemize}
\item Semantic ontology for descriptions of QoS properties including quality definitions, measures, calculations, etc;
\item Quality enable repository, which stores services descriptions, enhanced with QoS;
\item Appropriate mechanisms for collecting reliable QoS information from providers claims, users feedback, agents or brokers, etc.;
\item Matching and ranking algorithms returning web service description that fits best to user requirements;
\item Semantic Ontologies for QoS aware web service description.
\end{itemize}
\begin{figure}[!ht]
	\centering
		\includegraphics[scale=0.7]{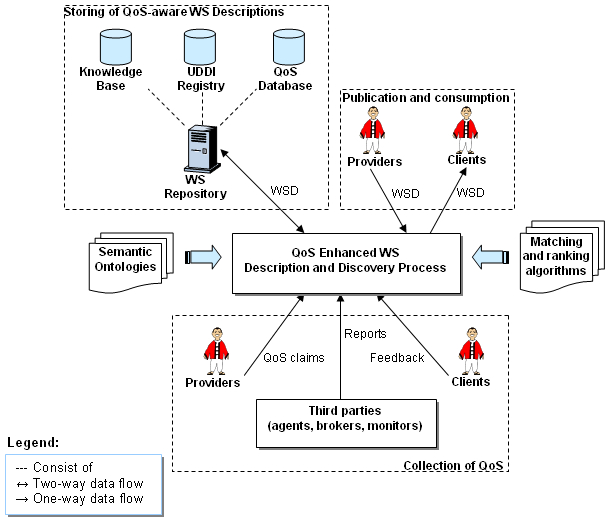}
	\caption{Components of Web service Description and Discovery Architecture.}
	\label{fig:Architecture}
\end{figure}
The component named 'Storing of QoS-aware WS Descriptions' points to the possibilities for storage of information about functional and non-functional web service properties. WS Repository has optional relations with UDDI registry, QoS Database and Knowledge base, which means that it could use a combination of them to maintain web service descriptions (WSD). The two-way relation of this component with 'Semantic WS Description and Discovery Process' shows that it could not only store, but also update information about web service properties.\\
The 'Publication and consumption' component represents the participants in the web service publication and the consumption process, respectively providers and clients. Clients receive WSD corresponding to their requirements after the execution of appropriate matching and ranking algorithms. Providers use the suitable ontology language to create WSD.\\
The 'Collection of QoS' component illustrates different ways for gathering in order to enhance the discovery process with semantic meaning. The participants in that process are providers claiming QoS, clients giving QoS feedback, and third parties such as agents, brokers or monitors providing QoS reports based on testing and monitoring.\\
The components described above are a base for creation of pattern for design of QoS enhanced web service description and discovery process. It is shown on Figure 2 as an use case diagram.\\
\begin{figure}[!ht]
	\centering
		\includegraphics[scale=0.75]{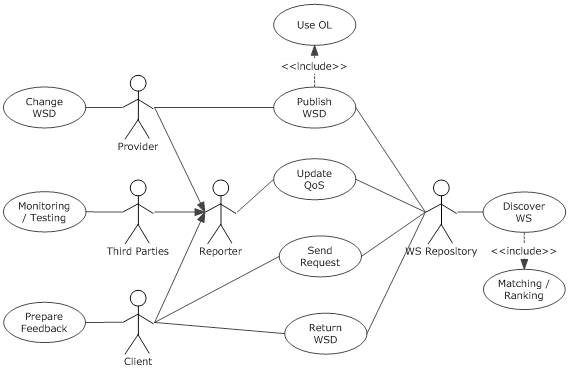}
	\caption{Use Case Diagram of Pattern for QoS Enhanced Web Service Description and Discovery.}
	\label{fig:UseCase}
\end{figure}
Our pattern presents the QoS enhanced web service description and discovery process on an abstract level. It determines the participants in the process and their respective roles and the main steps, leading to the achievement of the final goal for the best web service selection. The 'Reporter' actor corresponds to the 'Collection of QoS' component from Figure 1. It integrates the activities of the providers, clients and third parties that are relevant to updating of QoS information. We choose use case diagrams as a tool for the presentation of our pattern because they are understandable for most of the web service designers and developers. Each part of the pattern could be implemented in a different way depending on the concrete context. For instance, the type of semantic ontology defines the format of the providers' offers and the clients' requests.

\section{Conclusions}
In this paper principal current approaches for web service description and discovery taking into account QoS properties have been presented. They were examined according to several aspects providing the essential knowledge about the manner in which QoS information can be implicated in the process of web service description and discovery. These aspects include ontologies and ontology languages for description of functional and QoS characteristics of web services; repository or database for storing web services descriptions; classifications of QoS attributes; descriptions of matching and ranking algorithms; collection of QoS information and trustworthy; programming technologies for implementation of proposed architectures. As a result of our research we support developers to combine and implement the web service description and discovery approaches that will ensure best quality and performance of service oriented architectures.\\
In the future, we plan to propose a framework for QoS enabled web service description, publication and discovery, taking into account the benefits of current approaches. What is generally missing from these approaches, except~\cite{1} and~\cite{2}, is reputation-based trust mechanism for web service selection. The percentage of trusted providers and clients and the rate of cheating users have an effect on the quality of selection and ranking results. That is why further work should be done on the development of matchmaking and ranking algorithms based on providers QoS claims, consumers QoS requirements and third party QoS statistics. A challenge in that direction should be to find the best way for monitoring of QoS properties and precisely measuring their values. The next step would be to research in details existing ontology languages for semantic web service description and based on their features to use the most appropriate of them in our future framework. It would be beneficial to extend some of these languages in order to describe QoS attributes more precisely. Finally, we will focus on the design of a repository for storing QoS-aware web service descriptions that could be integrated with the UDDI registry. Special attention will be given to the research results that are presented during NFPSLA-SOC Workshops that are not covered in this paper~\cite{27,28}.

\end{document}